# Canonical Cortical Field Theories

Gerald K. Cooray[1,2], Vernon Cooray[3] and Karl Friston[4]

[1] GOS-UCL Institute of Child Health, University College London, London, UK

[2] Karolinska Institutet, Stockholm, Sweden

[3] Department of Electrical Engineering, Uppsala University, Uppsala, Sweden

[4] The Wellcome Centre for Human Neuroimaging, Queen Square Institute of Neurology, University College London, London, UK

Email: gerald.cooray@ki.se

Abstract

We characterise the dynamics of neuronal activity, in terms of field theory, using neural units placed on a 2D-lattice modelling the cortical surface. The electrical activity of neuronal units was analysed with the aim of deriving a neural field model with a simple functional form that still able to predict—or reproduce—empirical findings. Each neural unit was modelled using a neural mass and the accompanying field theory was derived in the continuum limit. The field theory comprised coupled (real) Klein-Gordon fields, where predictions of the model fall within the range of experimental findings. These predictions included the frequency spectrum of electric activity measured from the cortex, which was derived using an equipartition of energy over eigenfunctions of the neural fields. Moreover, the neural field model was invariant, within a set of parameters, to the dynamical system used to model each neuronal mass. Specifically, topologically equivalent dynamical systems resulted in the same neural field model when connected in a lattice; indicating that the fields derived could be read as a canonical cortical field theory. We specifically investigated non-dispersive fields that provide a structure for the coding (or representation) of afferent information. Further elaboration of the ensuing neural field theory—including the effect of dispersive forces—could be of importance in the understanding of the cortical processing of information.

# 1 Introduction

The experimental and theoretical studies of Hodgkin and Huxley resulted in a foundational attempt to model neuronal dynamics (Hodgkin et al 1954). The Hodgkin-Huxley model determines the pointwise activity of a membrane. The model comprises a 4-dimensional, nonlinear ordinary differential equation describing sodium and potassium dynamics. It accurately models the membrane behaviour of voltage-gated ion channels. This model can be extended to include the effect of different types of ion-channels and to include the spatial effects of an extended membrane. Modelling the dynamics of several neuronal cells results in a set of nonlinear differential equations which are complicated to analyse, and there have been several attempts to simplify the implicit model. These have included different types of neural mass approximations. One of the simplest neural mass models is the integrate and fire neuron (or the leaky integrate and fire neuron), where the neuron fires when the membrane potential reaches a threshold (Lapiquie 1907, Brunel et al 2007).

The integrate and fire neuron is still one of the most common approximations in network models of neural tissue, mainly due to its simplicity. The interaction between neurons in a network or sheet is often approximated using a sigmoid function, mapping presynaptic potentials to postsynaptic currents, and a weight-function to parametrise connection strength. Each neuron receives input from a large number of neighbouring neurons and the total effect can be approximated using a sum or integral. The above steps result in an integro-differential equation for the neuron determining the dynamics. This can be analysed in the continuum limit resulting in neural field equations which have been analysed by (Wilson and Cowan 1973, Amari 1977, Bresloff 2012, Coombes 2005, and Ermentrout and Cowan 1979) to mention a few. The above models approximate the quasi-microscopic dynamics of neural tissue well and several predictions of the ensuing field theory have been validated experimentally (Chervin *et al*. 1988; Golomb & Amitai 1997, Wu *et al* 1999, Miles *et al*. 1995).

The mathematical structure of the models described above is based on a one-dimensional partial differential equation or of two dimensions, if the unit that is modelled consists of an excitatory and inhibitory neuron. The second coupled equation has also been implemented by introducing a nonlinearity into synaptic depression. In contrast, the Hodgkin-Huxley model is 4 dimensional, which makes non-numerical analysis of the continuum model very difficult. The FitzHugh Nagumo model is a 2-dimensional approximation of the former (FitzHugh 1961, Nagumo et al 1962) with 2 variables representing a fast and slow parameter, where the dynamics resemble that of the Hodgkin-Huxley model. The FitzHugh model has been implemented in the continuum limit of a cortical sheet; with similar predictions to that made by neural fields based on integrate and fire neurons (Ermentrout et al 1984). This suggests an underlying universality—in the activity modelled by neural fields—where the exact structure of the underlying model is of less importance.

Even though it is understood that cortical activity progresses over a surface, experimental measurement of that activity is always performed at specific points. Modelling this sparse sampling of cortical activity has led to a vast literature on pointwise neural models, often in finite networks. The neural mass models constructed are most often coupled 2nd order ODEs where the basic neuronal unit is 2 dimensional and represents a population of a specific neuronal cell-type. These models have provided compelling models of electrographic and magnetographic data from the human and animal brain; including activity seen in stimulation related changes of cortical activity and the spreading of epileptic seizures (Lopes da Silva 1997, Wendling et al 2000, Kiebel et al 2008, Jirsa et al 2014, Jirsa et al 2018). In this paper our goal will be to derive a neural field, such that it retains a high degree of mathematical simplicity (i.e., symmetry or invariance structures) and yet is sufficiently expressive to model the features of cortical activity seen empirically. Crucially, under this formalism, it can be shown that the neural field dynamics is independent of the specific type of underlying neural mass dynamic. The neural field we consider is governed by a set of coupled wave-equations of Klein Gordon type. To establish predictive validity of the ensuing field model, its behaviour was compared with experimental findings. To establish the construct validity—in relation to computational constructs, we briefly consider a scheme for information processing and representation of afferent input by Klein Gordon type neural fields.

## 2   The neural field model

The cortex comprises several layers of neuronal populations with distinct dynamic properties and reciprocal connections. There are strong (intrinsic) connections between cortical layers running orthogonal to the surface of the cortex and weaker tangential (horizontal) connections along the surface; allowing for the definition of a cortical column with intrinsic (inter-and intra-laminar) processing between layers interconnected with weaker connections between columns (Douglas et al 1989). We consider a standard cortical column where the lateral tangential connections play a nontrivial role. The extent of lateral connections is approximated as part of a square lattice with a node and 4 nearest neighbours. In the continuum limit, the resulting neural field will be a N-vector field (each layer representing 1 dimension of the vector). The connections between different cortical layers will be modelled using different types of synaptic kernels. We will use a distinct kernel for potential-to-current (S-type) and current-to-current (P-type) connections. The S-type kernel models axonal-to-dendritic interaction through chemical synapses and is often parameterised by a sigmoidal curve (tanh was used in this study). The P-type kernel models firing rate adaptability via dendritic interaction, including the self-interaction of a neuronal population (usually caused by membrane current-leakage). Please see Fig 1 for a schematic of the lattice model.

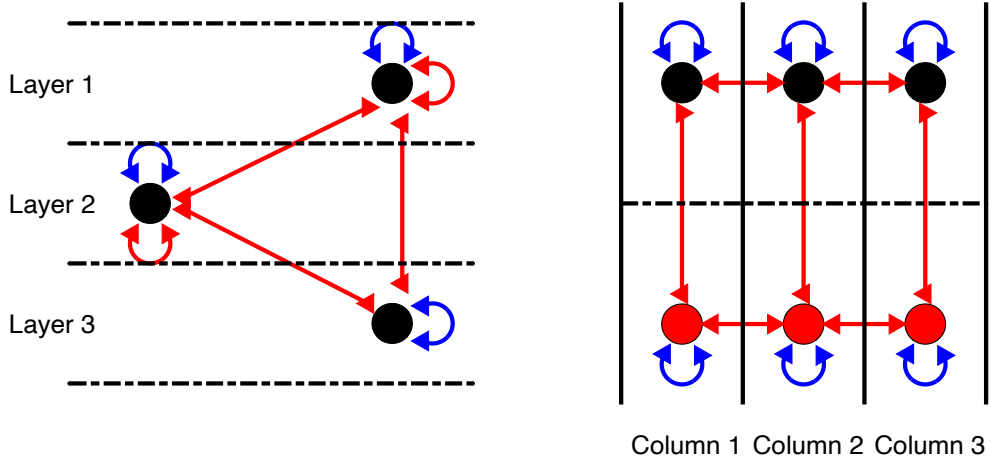

**Fig 1.** Schematic of neuronal populations and connections within a cortical column (left panel) and between cortical columns (right panel). Connectivity between neuronal populations are shown as arrows, where a kernel mapping potentials-to-currents are drawn in red and those mapping currents-to-currents in blue.

We model the activity of a cortical column using coupled 1st order ODEs giving us a description of a point of the cortex, i.e., a neural mass model. The following nomenclature and definitions will be used in the following sections.

$p^i_\mu$ – current variable of the $i^{th}$ population (cortical layer i) of the $\mu^{th}$ cortical column

$q^i_\eta$ – potential variable of the $i^{th}$ population (cortical layer i) of the $\mu^{th}$ cortical column

$\omega^i$ – angular velocity of the $i^{th}$ neuronal population (layer i) of a cortical column

S – A synaptic kernel mapping potentials to currents

P – A synaptic kernel mapping currents to currents

$s_{\mu\eta}^{ij}$ -- Connectivity between population i in cortical column $\mu$ and population j in cortical column $\eta$ via synapse of type S.

$p_{\mu\eta}^{ij}$ -- Connectivity between population i in cortical column $\mu$ and population j in cortical column $\eta$ via synapse of type P.

A cortical column is modelled by the following coupled 1st order ODE equations; the effect of the different layers is included in the sums (roman indexes) as are the effects of adjacent cortical columns (Greek indexes). The P coupling is assumed to be mediated through self-interactions (i.e. current leakage across the membrane)

$$\dot{p}_\mu^i = -\omega_i^2 q_\mu^i + \omega_i^2 \sum_j s_{\mu\mu}^{ij} S(q_\mu^j) + \omega_i^2 \sum_\eta s_{\mu\eta}^{ii} S(q_\eta^i) + \omega_i p_{\mu\mu}^{ii} P\left(\frac{p_\mu^i}{\omega_i}\right)$$

$$\dot{q}_\mu^i = p_\mu^i$$

Connections between columns is layer wise, but connections within a column connect different layers and include self-connection terms. The first sum runs over intrinsic connections (roman indexes, within column connections) and the second sum over extrinsic connections (Greek indexes, between columns connections). We start our analysis with near zero activity, where both the S and P kernels are linear to first order.

The above equation can be rewritten as follows using a linear approximation to the kernels.

$$\dot{p}_\mu^i = -\left(\omega_i^2 - \sum_\eta \omega_i^2 s_{\mu\eta}^{ii}\right) q_\mu^i + \sum_j \omega_i^2 s_{\mu\mu}^{ij} q_\mu^j + \sum_\eta \omega_i^2 s_{\mu\eta}^{ii}(q_\eta^i - q_\mu^i) - \omega_i p_{\mu\mu}^{ii} p_\mu^i$$

Assuming that the extrinsic connections are symmetrical over the lattice and equal for each layer (i.e., A) we get the following,

$$\dot{p}_\mu^i = -\left(\omega_i^2 - \sum_\eta \omega_i^2 s_{\mu\eta}^{ii}\right) q_\mu^i + \sum_j \omega_i^2 s_{\mu\mu}^{ij} q_\mu^j + A \sum_\eta (q_\eta^i - q_\mu^i) - \omega_i p_{\mu\mu}^{ii} p_\mu^i$$

Furthermore, we will assume that the intrinsic connections are real and symmetric (giving us a Hermitian "inter-layer mixing" operator).

$$H_j^i = -\omega_i^2 s_{\mu\mu}^{ij} = -\omega_j^2 s_{\mu\mu}^{ji}$$

$$H_i^i = \omega_i^2 - \sum \omega_i^2 s_{\mu\eta}^{ii}$$

Re-writing the coupled equations using matrix notation (in bold) gives the following reduced set of dynamical equations.

$$\dot{p}_\mu = -\mathbf{H} q_\mu + A \sum_\eta (q_\eta - q_\mu) - \mathbf{B} p_\mu$$

In the continuum limit, we have the following equation (with suitable re-definition of the spatial scale for matrices and scalars).

$$\dot{p} = -\mathbf{H}q + A\Delta q - \mathbf{B}p$$

If B is small enough (to be neglected), we have the following equation (after eliminating **p**).

$$\ddot{q} = -\mathbf{H}q + A\Delta q$$

Diagonalizing **H** (given by a diagonal matrix, D)—using an orthonormal basis—we get the following equation, comprising N-uncoupled wave equations (i.e., N real Klein-Gordon fields).

$$\ddot{q} - A\Delta q + \mathbf{D}q = 0$$

The propagation speed and mass-term of the waves is given by the following expressions,

$$c^2 = A$$

$$m_i^2 = \frac{\{\mathbf{D}\}_{ii}}{A^2}$$

Positive mass terms will result if the **H**-matrix is dominated by the first term in Eq. 2.1. With this basic approximation in place, one can now ask how it depends upon different functional forms for the neural mass model. We will see that the above formalism remains unchanged under some minimal assumptions about the different neural mass dynamics one might entertain.

## 3  INVARIANT NEURAL FIELDS WITH TOPOLOGICALLY EQUIVALENT PHASE SPACE DYNAMICS.

In the previous section, field dynamics was derived from a lattice approximation of the cortical surface where each node was modelled by a neural mass. In this section, we examine the dependency of the ensuing neural field dynamics on the specific neural mass used to model each unit of the lattice. Each neural mass is characterized as a dynamical system—in terms of its phase portrait—where systems that are topologically equivalent have equivalent phase portraits. A diffeomorphism between the two phase portraits is sufficient for topological equivalence (Arnold 2013).

The derivation of the field dynamics in section 2 was considered for small oscillations of the neuronal unit around a stable point, so we only need to study a small region of the phase space surrounding the stable points for distinct dynamical systems modelling the neuronal mass. We now provide the analysis for one layer of the cortex (i.e. a 2 dimensional phase portrait). Assuming there is a local diffeomorphism between the phase portraits of the different neural mass models, the transformation of variables between the two systems will be given by the following,

$$F: \mathbb{R}^2 \to \mathbb{R}^2$$

$$\mathbf{J}: T_{(0,0)}\mathbb{R}^2 \to T_{(Q,P)}\mathbb{R}^2$$

$$\begin{pmatrix} q' \\ p' \end{pmatrix} = \begin{pmatrix} Q \\ P \end{pmatrix} + \mathbf{J} \begin{pmatrix} q \\ p \end{pmatrix}$$

**J** is the Jacobian for the mapping *F* between the two phase portraits and (*Q*,*P*) are the coordinates that the stationary state (0,0) maps to. The neural field derived in section 2 are given by the following equation,

$$\begin{pmatrix} \ddot{q} \\ \ddot{p} \end{pmatrix} = \begin{pmatrix} -D + A\Delta & 0 \\ 0 & -D + A\Delta \end{pmatrix} \begin{pmatrix} q \\ p \end{pmatrix} \quad \text{Eq. 3.1}$$

Which, in the second system, are given by the following equation,

$$\frac{d^2}{dt^2} \begin{pmatrix} q' - Q \\ p' - P \end{pmatrix} = \mathbf{J} \begin{pmatrix} -a + b\Delta & 0 \\ 0 & -a + b\Delta \end{pmatrix} \mathbf{J}^{-1} \begin{pmatrix} q' - Q \\ p' - P \end{pmatrix} \quad \text{Eq. 3.2}$$

One can show as follows that the above equation (Eq. 3.2) is identical to the neural field equation derived using the first dynamical system (Eq. 3.1).

$$\frac{d^2}{dt^2} \begin{pmatrix} q' - Q \\ p' - P \end{pmatrix} = \mathbf{J} \begin{pmatrix} -a + b\Delta & 0 \\ 0 & -a + b\Delta \end{pmatrix} \mathbf{J}^{-1} \begin{pmatrix} q' - Q \\ p' - P \end{pmatrix} = \begin{pmatrix} -a + b\Delta & 0 \\ 0 & -a + b\Delta \end{pmatrix} \mathbf{J}\mathbf{J}^{-1} \begin{pmatrix} q' - Q \\ p' - P \end{pmatrix}$$

$$= \begin{pmatrix} -a + b\Delta & 0 \\ 0 & -a + b\Delta \end{pmatrix} \begin{pmatrix} q' - Q \\ p' - P \end{pmatrix}$$

These derivations show that the field equations are invariant to the dynamical system used to model the neuronal unit (assuming a system with a 2 dimensional phase portrait). In other words, the properties derived for the neural field of the previous section are identical for topologically equivalent phase space dynamics (this can be shown not to hold in general for higher dimensions). We now turn to the dynamical features that this model exhibits.

# 4 High amplitude semi-stationary sets and their propagation.

In previous work (Cooray et al 2023a, Cooray 2023b) we have shown that semi-stable limit cycles of a single cortical column occur when the current-to-current coupling kernel has several zero points. After performing an adiabatic average over the phase portrait, we derived the stable points for the amplitude of the activity, showing the presence of stable limit cycles. These high amplitude states could represent rhythmic activity seen in recordings from the human cortex including both healthy and abnormal activity. The speed of propagation, for a spreading limit cycle front—across the cortical surface—can be derived along a line. As an initial configuration, we will assume that the left half of the one dimensional lattice is in a high amplitude oscillatory state (i.e. a limit cycle activity). The other half of the lattice terms are assumed to have low amplitude, near zero, dynamics. We assume only nearest neighbour interactions. The dynamical equation is as follows (where R and φ are amplitude and phase variables along the lattice and the subscript indicates the position along the lattice increasing from left to right),

$$\dot{R}_0 - i\dot{\varphi}_0 R_0 = i\omega s_{0-1}(1 - i(\varphi_{-1} - \varphi_0))R_{-1} + i\omega s_{0,1}(1 - i(\varphi_1 - \varphi_0))R_1 - \omega p_{0,0} B' R_0$$

$$\dot{R}_0 - i\dot{\varphi}_0 R_0 \approx i\omega s_{0-1}(1 - i(\varphi_{-1} - \varphi_0))R_{-1}$$

$$\dot{R}_0 = (\omega s_{0-1}(\varphi_{-1} - \varphi_0))R_{-1}$$

$$\dot{R}_0 = -\omega s_{0-1} R_{-1} \frac{d\varphi}{dx} \Delta x$$

The time taken for activity to move from an amplitude near 0 to R$_{final}$ will be,

$$\frac{R_{final} dx}{-\omega s_{0-1} R_{-1} d\varphi \Delta x} = \Delta t$$

The front would have then moved from position -1 to 0. The speed of propagation of the wave front would then be,

$$\frac{\Delta x}{\Delta t} = -\omega s_{0-1} \frac{d\varphi}{dx}$$

The above equation uses an arbitrary length scale. Taking into consideration the size of a neuronal unit, d, the speed will be given by the following expression,

$$\frac{dx}{dt} = -d\omega s_{0-1} \frac{d\varphi}{dx}$$

One can see that for a positive gradient of the phase lag, the wave will move from left to right ($s_{0-1}>0$). The effect of near zero activity on high amplitude activity is negligible as can be seen from the above equations. Characterising dynamics in terms of travelling waves in this way affords the opportunity to compare predictions of the neural field approximation to empirical results, as briefly reviewed in the next section.

# 5 EXPERIMENTAL SUPPORT FOR DERIVED QUANTITIES

Experimental studies have detected travelling waves in cortical tissue with speeds ranging from 0.001 to 30 ms$^{-1}$ (Muller et al 2018, Xu et al 2007, Sanchez-Vives MV et al 2000, Wester et al 2012, Muller et al 2012). The speed of propagation of the waves derived in section 2 and 4 can be estimated using a few steps. The variables required for this estimation are listed below, with the range of empirical values.

ω: natural frequency of the neuronal subpopulations in a cortical column (approx. 10-1000 rad/s, (Steriade et al, 1993; Simon et al, 2014)

g: connectivity between neuronal subpopulations (1-10) (Fastenrath et al 2009).

Σ: number of neuronal units affecting a given subpopulation (square lattice, with 5 units in sum)

W: weighting function with nearest neighbour-connectivity (Amari et al, 1977).

d: size of cortical column (approx. 0.0001-0.001m) (Dalva et al, 1997, Mountcastle et al, 1955))

The weighting function (W) of neuronal interactions has been determined in the primary visual cortex to span approximately 1 cortical column. Translating this to the square lattice approximation used above, we arrive at nearest neighbour excitation and a very quick drop in synaptic gain at larger distances. In the continuum limit, this is approximated with the diffusion operator and a self-interaction term (note that the equations are only correct to order of magnitude).

$$a + b\Delta \sim \omega^2 \sum_{(neighbors)}^{N} g \sim \omega^2 |g| d^2 \sum_{(neighbors)}^{N} \frac{g}{|g| d^2} \sim c^2 \Delta$$

In the last step, we have included the square of the dimension of the internodal spacing of the lattice which we are replacing with the Laplacian operator. The term in the sum is approximated with the Laplacian operator and the preceding term determines the conduction speed of the travelling waves.

$$c \sim 0.001 - 1 \text{ms}^{-1}$$

This value is in the range of values determined experimentally for travelling waves in cortical tissue. The derivations in section 2 and 4 also estimated a mass term for the travelling waves which would be of the order $c^{-2}$. In section 3, we determined the spread of a limit cycle front which can be estimated to the following value (using d=0.001 m and ω=100),

$$\frac{dx}{dt} = 0.001 * 100 * \frac{d\varphi}{dx}$$

The rate of change of angular frequency is estimated to about 1 radian per 0.01m, estimated form measurements from humans using intracranial electrodes (Kramer et al 2005). This will give a speed of about 10ms$^{-1}$, which again falls within the experimental range (Kramer et al 2007). The frequency of the travelling waves—estimated using the free solution to the Klein Gordon equation— section 2) can be determined by the expression for the total energy of the wave and will be given by the following expression,

$$E^2 = m^2 c^4 + p^2 c^2 = 4\pi^2 f^2 \qquad \text{Eq. 5.1}$$

The spatial resolution of the waves will be limited by the size of a cortical column. This gives a range of momenta (p) for the waves from about 1 - 1000 m$^{-1}$. The corresponding frequency range will be given by,

$$f = 0.1 - 200 \text{ Hz}$$

This range fits well with frequencies observed in electrical recordings from cortical tissue in both invasive and non-invasive sampling. Moreover, we can estimate the frequency distribution of EEG sampled on the cortex by assuming equidistribution of energy over the different frequency components of the spectrum. The energy density of the waves is given by the following expression, (where ρ denotes the density of frequency states and using Eq. 5.1),

$$\rho(f)E(f) = \rho(f)2\pi f$$

From the equipartition of states we assume this to be independent on the states giving the following,

$$\rho(f) = \frac{A}{2\pi f}$$

This estimate is in good concordance with experimental data showing an inverse frequency distribution for the frequency power spectrum. In summary, the neural field approximation yields internally consistent estimates of travelling waves that fit comfortably with empirical observations of endogenous dynamics; at least two within an order of magnitude. We next consider exogenous or external perturbations in terms of neural field responses to, e.g., sensory afferents or (extrinsic) projections from other neural fields.

# 6 INTERACTION WITH EXTERNAL INPUT

So far we have studied the endogenous (i.e., autonomous) dynamics of the cortical lattice in the absence of external input. The dynamics will modify accordingly when each cortical column is connected to an external input source (we will assume the one-layer model governed by the Klein Gordon wave equation, section 2). The requisite equation of motion (simplified to remove constants) is given by the following expression,

$$\ddot{q} - \Delta q + q = I(t, x)$$

Eq. 6.1

*I(t,x)* denotes input to the cortical column and *x* is the position vector on the cortical surface. *I(x,t)* will be approximated by an initial value condition.

$$q(x, 0) = I(x)$$
$$\dot{q}(x, 0) = 0$$

The spatial Fourier transform of *I(x)* is given by the following expression (parameter k is a vector, momentum vector),

$$\hat{I}(k) = \int I(x) e^{-ix.k} dx$$

The solution to Eq. 6.1 is given by the following,

$$q(t,x) = \frac{1}{2\pi} \int \frac{\hat{I}(k) e^{ix.k}}{2} \left( e^{it\sqrt{k^2+1}} + e^{-it\sqrt{k^2+1}} \right) dk \qquad \text{Eq. 6.2}$$

Eq 6.2 represents the cortical response and implicit representation of the exogenous input, *I(x)*. A basis (*e(k)*) for the implicit encoding of inputs can be expressed as follows,

$$e(k) = \frac{1}{4\pi} e^{ix.k} \left( e^{it\sqrt{k^2+1}} + e^{-it\sqrt{k^2+1}} \right)$$

$$f = \int \hat{I}(k) e(k) \, dk$$

Or

$$f = \sum \hat{I}_k e_k$$

Depending on the size of the cortical column, one of the above expressions can be used (if the cortical sheet is modelled using an infinite sized domain the integral expression will be used; otherwise, the discrete sum is valid). The vector Î(k) or $\hat{I}_k$ can be seen as a coordinate expression for the encoded input data. The coordinates can be used to define a statistical manifold. However, this would require that a metric be defined on the tangent bundle of the statistical manifold, which we will not pursue in this paper.

# 7 DISCUSSION

In this paper, we have derived neural field equations for (patches of) the cortex using a lattice approximation of the cortical sheet with neural mass models placed at each node and nearest neighbour connections. When cortical layers are decoupled, the resulting field equations were shown to be invariant on the type of neural mass model used on the lattice. More specifically, we derived a set of real Klein-Gordon fields, where the individual fields had un-equal mass terms for interacting cortical layers and equal mass terms for decoupled layers. Due to the simplicity of the pursuing neural fields, including invariance properties of the fields, we considered them to be of a canonical form, representing a canonical cortical field.

The intrinsic and extrinsic connectivity determined the type of neural fields generated by the lattice structure. The balance between intrinsic self-connection and extrinsic connections defined the balance between the Laplacian operator and mass term in the resulting Klein-Gordon field. We have, so far, investigated symmetric connections between layers resulting in a set of uncoupled (approximately) non-dispersive fields. These fields only showed attenuation of activity due to an underlying current leakage across cell membranes of the underlying neuronal units, which we disregarded in the analysis (and should be an acceptable approximation in a time frame of short enough duration). However, including non-symmetric connections would result in partially dispersive fields. This is most easily seen by writing the cortical connectivity matrix as a sum of a real Hermitian and skew Hermitian matrix, where the dynamics generated by the skew Hermitian matrix would be dispersive. A system with only non-dispersive fields allows for storage of afferent information. We considered a system where exogenous input was shown to map to a function space defined by the canonical cortical field. One can then read this mapping in terms of an encoding of afferent input. It could be presumed that any processing of the encoded input would be mediated by the interaction between the dynamics generated by Hermitian and skew Hermitian connectivity matrices. The space onto which the input is mapped—together with a metric on that space (i.e., the Fisher Metric)—could be treated as a statistical manifold (Ay et al 2017). However, as the implications of skew symmetric matrices and the subsequent processing of data were not examined, the introduction of a metric is still speculative and not further elaborated here.

The derivation of the neural field model assumes certain connections weights between cortical columns. Connectivity weights were assumed to be positive within nearest neighbours and then quickly decrease with distance. This shape of the waveform is a lattice approximation of the Mexican hat distribution for weights used by previous authors in deriving neural fields. This resembles the weights seen experimentally in animal study of cortical tissue (Lund et al 2003). Analysis of the Mexican hat distribution of weights results in a power series of differential forms where the lowest order approximation corresponds to a diffusion equation (Amari 1977, Coombes 2005, Bresloff 2012, Dalva et al 1997). In the lattice model of the cortical sheet, this distribution would be equivalent to the discrete Laplacian operator and a self-connection term. This was the assumption that allowed for the conversion of the point theory of cortical columns to a field theory over the cortical sheet.

In vivo and vitro studies of cortical tissue have shown a wide variation in conduction speeds of waves from very slow speeds of about 0.001 ms$^{-1}$ to maximal conduction speeds in the range of axonal conduction of about 10-30ms$^{-1}$. The propagation speeds of canonical cortical fields depend partly on the size of a cortical column and the underlying characteristic frequency of the neuronal populations, which were determined from the literature and were predicted in the lower range of the experimentally observed values (Dalva et al 1997, Domich et al 1986, Bragin et al 1999, Staba et al 2002). The prediction of travelling waves fits well with experimental data and also predicts the large variability in conduction speeds. In vivo data from humans have shown a variable prediction for the propagation of seizure activity across the cortical surface (Hall et al 2013). With our models, propagation speed for high amplitude activity (a proxy for seizure activity)—in a background of near zero activity—was predicted and shown to be variable as depends dynamically on the underlying phase gradient. The frequency content and distribution of electrographic activity has been measured consistently using both non-invasive and invasive methods in humans, where the main spectral content shows broad band activity, approximately 1-200Hz (Niedermeyer et al 2005). The frequency bandwidth falls in the range of the theoretical predictions of the neural field model, where the spatial extent and its graininess determines the upper and lower bandwidth limits. The lower limit was estimated to the size of a cortical column (0.1-1mm) and the full sheet was approximated to 0.1-1m$^2$. Finally, the frequency distribution was estimated using a simple equidistribution of energy over eigenstates of the neural field model. There does not seem to be a consensus in the literature on the reason for the inverse relationship with frequency in electrographic data, but several theories have been presented including phase stability, critical states and biophysical filtering (Thatcher et al, 2009, Bedard et al 2006).

In conclusion, this study has derived a cortical field theory based on a lattice architecture of neural mass models over the cortical surface. The construction allowed the leverage of certain invariance (i.e., symmetry) properties of the neural mass, resulting in a cortical field with partially rotational invariance. The main predictions of the model fall within the realm of experimental findings and are similar to other neural field models. In contrast to numerical characterisations of neural field models, we have focused on a highly symmetric model, the real Klein-Gordon field with well-known solutions. The assumptions and approximations performed in the derivations could be criticized for being overly simplistic and constrained. However, the goal was to present a field model retaining some of the important characteristics of neural fields yet being simple enough for analytic investigation. Moreover, the neural field shows invariance, in relation to the neural mass models used to derive it, speaking to its canonical properties. Further elaboration of the neural field as discussed above including information coding and processing properties could be of interest within the field of cognitive neuronal computation.

# 8 FUNDING

KF is supported by funding for the Wellcome Centre for Human Neuroimaging (Ref: 205103/Z/16/Z), a Canada-UK Artificial Intelligence Initiative (Ref: ES/T01279X/1) and the European Union's Horizon 2020 Framework Programme for Research and Innovation under the Specific Grant Agreement No. 945539 (Human Brain Project SGA3).